\documentclass{elsart}
\usepackage{epsfig}
\usepackage{amssymb}

\begin{document}
\begin{frontmatter}

\title{ A  closer look at the indications of $q$-generalized Central Limit Theorem behavior   in  quasi-stationary states of the HMF model 
} 

\author
{Alessandro Pluchino} \ead{alessandro.pluchino@ct.infn.it}
\address{Dipartimento di Fisica e Astronomia,  Universit\'a di Catania,\\
and INFN sezione di Catania,  Via S. Sofia 64, I-95123 Catania, Italy} 

\author
{Andrea Rapisarda} \ead{andrea.rapisarda@ct.infn.it}
\address{Dipartimento di Fisica e Astronomia,  Universit\'a di Catania,\\
and INFN sezione di Catania,  Via S. Sofia 64, I-95123 Catania, Italy} 

\author
{Constantino Tsallis}  \ead{tsallis@cbpf.br}
\address {Centro Brasileiro de Pesquisas Fisicas, Rua Xavier Sigaud 150, 22290-180 Rio de Janeiro-RJ, Brazil \\
and \\
Santa Fe Institute, 1399 Hyde Park Road, NM 87501, USA}

%\date{today}
%
%\pacs{05.45.-a, 05.65.+b, 89.75.-k}
%
%\maketitle

\begin{abstract}

% Text of abstract

We give a  closer look  at  the   Central Limit Theorem  (CLT) behavior  in  quasi-stationary states  of  the  Hamiltonian Mean Field model, a paradigmatic one  for long-range-interacting classical many-body systems. We present  new calculations  which  show  that, following their time evolution,  we can observe and classify  three kinds  of long-standing quasi-stationary states (QSS) with different  correlations. The  frequency of occurrence of each  class  depends on the size of the system.  The  different   microsocopic nature of the QSS    leads  to different dynamical correlations and therefore to different    results  for the observed CLT behavior.

\end{abstract}

\end{frontmatter}

\section{Introduction}

Very recently  there has been  a lot of interest in   generalizations  of  the  Central Limit Theorem (CLT) \cite{clt0,umarov,barkai} and on  their   possible (strict or numerically approximate) application   to   systems  with long range correlations \cite{luis,hil}, at the edge-of-chaos \cite{ucc}, nonlinear dynamical systems  the maximal   Lyapunov exponent  of which is either  exactly zero or  tends to vanish in the thermodynamic limit (increasingly large systems) \cite{anteneodo}, hindering  this way  mixing  and thus the application  of standard statistical mechanics.  A possible application of  nonextensive statistical mechanics  \cite{cost1,clt1} 
has  been  advocated  in these cases.
Along this  line   we discuss  in the present  paper  a   detailed  study of   a  paradigmatic \textit{toy model}   for  long-range  interacting  Hamiltonian   systems \cite{hmf,liap,epn-rap,chavanis,fulvio1,ruffo,reply07}, i.e.  the     Hamiltonian  Mean Field (HMF) model  which has been intensively   studied  in the last years.   
In a recent article \cite{rapis-ctnext07},
we presented  molecular dynamics numerical results for the HMF model  showing  three kinds of quasi-stationary states (QSS)
starting from the same water-bag initial condition with unitary magnetization ($M_0=1$). 
 In the following  we present   how the applicability of the   standard or $q$-generalized CLT is influenced by  the different microscopic dynamics observed in the  three classes of QSS.  In general, averaging   over the threee  classes  can be misleading. Indeed, the frequency of appearence of each of these classes depends on the size of the system under investigation, and there is no clear evidence that  a predominant class exists.

\section{Quasi-stationary behavior in the HMF model}

The HMF model consists  of $N$ fully-coupled classical inertial
XY spins (rotors)
$\stackrel{\vector(1,0){8}}{s_i} = (cos~\theta_i,sin~\theta_i)~,~i=1,...,N$, 
with unitary module and mass \cite{hmf}. One can also think of 
these spins can  as rotating
 particles  on the  unit circle.
The  Hamiltonian is given by
\begin{equation}
\label{hamiltonian}
        H
= \sum_{i=1}^N  {{p_i}^2 \over 2} +
  { 1\over{2N}} \sum_{i,j=1}^N  [1-cos( \theta_i -\theta_j)]~~,
\label{eq.2}
\end{equation}
where ${\theta_i}$ ($0 < \theta_i \le 2 \pi$) is the angle
and $p_i$ the conjugate variable representing the rotational
velocity of spin $i$.
\\
At  equilibrium the model can be solved exactly and the solution 
predicts a second order phase transition from a high
temperature paramagnetic  phase to a low temperature
ferromagnetic  one \cite{hmf}. The transition occurs at the
 critical temperature $T_c=0.5$  which  corresponds to
a critical energy per particle $U_c = E_c /N =0.75$.
The order parameter  is the modulus of
the {\it average magnetization} per spin defined aa
$M = (1 / N) | \sum_{i=1}^N
\stackrel{\vector(1,0){8}}{s_i} | $.
Above $T_c$,  the spins  are  homogeneously distributed  on the circle so that   $M \sim 0$, while 
below $T_c$, most spins
are aligned, i.e. rotors are trapped in a single cluster, and $M \neq0$.
The out-of equilibrium dynamics of the  model presents   very interesting dynamical anomalies. For  
 energy densities  $U\in[0.5,0.75]$, special  classes of initial conditions such as those called
\textit{water-bag}, characterized by an initial magnetization $0 \le M_0 \le 1$  and uniform distribution of the momenta,
drive the system, after a violent relaxation,
towards  metastable QSS. The latter slowly decay towards equilibrium with a lifetime which diverges
like a power of the system size $N$ \cite{hmf}.
\begin{figure}   %[h,t]
\begin{center}
\epsfig{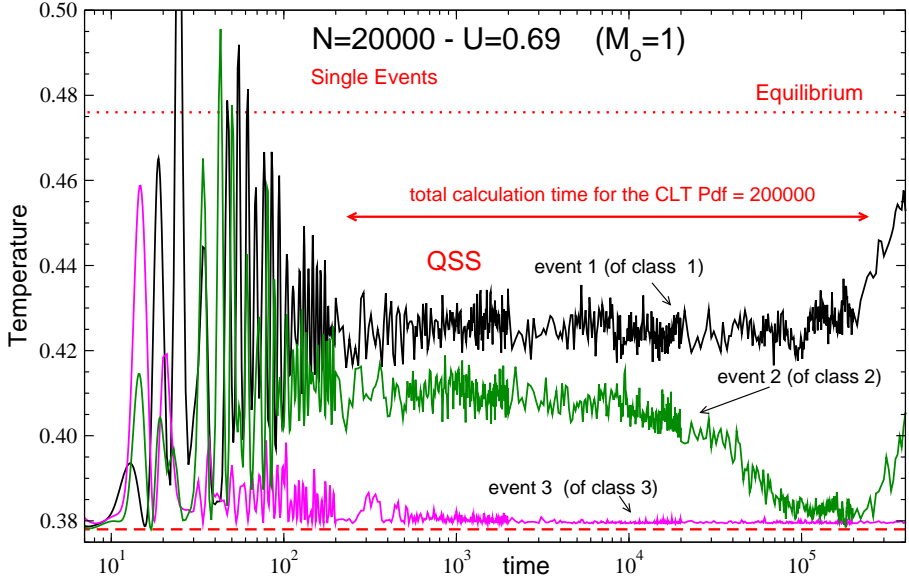}
\end{center}
\caption{ Time evolution of the temperature (calculated as twice the average kinetic energy per particle) for three  single events representative of the three different  classes observed at $U=0.69$ for initial magnetization $M_0=1$.  
The size of the system is $N=20000$. 
}
\end{figure}
Along the  QSS   regime, in the microcanonical ensemble,  the  dynamics exhibits    a glassy behavior, hierarchical structures, velocity correlations, aging, vanishing Lyapunov exponents, etc.,  and  the statistical description of the QSS is strongly dependent on the initial conditions \cite{liap,epn-rap,chavanis}. However,   even  for the   same type of initial conditions,  we have observed  different dynamical behaviors   \cite{rapis-ctnext07}.
In particular we have found  three  typical classes  of events, as illustrated in Fig. 1. Here  we plot  the  temperature of the system   (defined as twice the average kinetic  energy per particle) versus time. We  focused on a system with size $N=20000$,  the  energy per particle being $U=0.69$,  for which anomalies are more evident. Initial conditions were  chosen to be of the water-bag kind, with initial magnetization equal to unity ($M_0=1$). 
For details about the accuracy of the calculations and  the integration algorithm
adopted see Ref.\cite{epn-rap}.
Only  single  events  are plotted in Fig. 1, each one representative of a given class. 
One immediately realizes  that there are two extreme events (which we indicated  with 1 and 3) and an intermediate one (indicated  as 2) which stays close to event  1 at the beginning,  but  collapses  towards the plateau of the event 3 after some time.  
The relative frequency of  occurrence  of these three classes of events is illustrated in  Fig. 2 as   a function of the size $N$ of the system. A total of $20$  realizations for each $N$ was considered. 
It is important to note that the event of class 3 shown in Fig. 1, and all the events of class 3 in general, are very similar to the  QSS obtained  for  initial conditions  with  zero initial magnetization ($M_0=0$),  which are almost homogeneous and  have very small correlations \cite{epn-rap}. 
In this case   a  Lynden-Bell  kind of  approach, or one based on the Vlasov equation   has been   applied \cite{ruffo}.  
Therefore, due to the different nature of these QSS,  which have in general   different  microscopic correlations, 
 one could  expect   a    different     result  for  the   central limit theorem behavior shown in Refs.  \cite{reply07,epl07,canberra}.

\begin{figure}   %[h,t]
\begin{center}
\epsfig{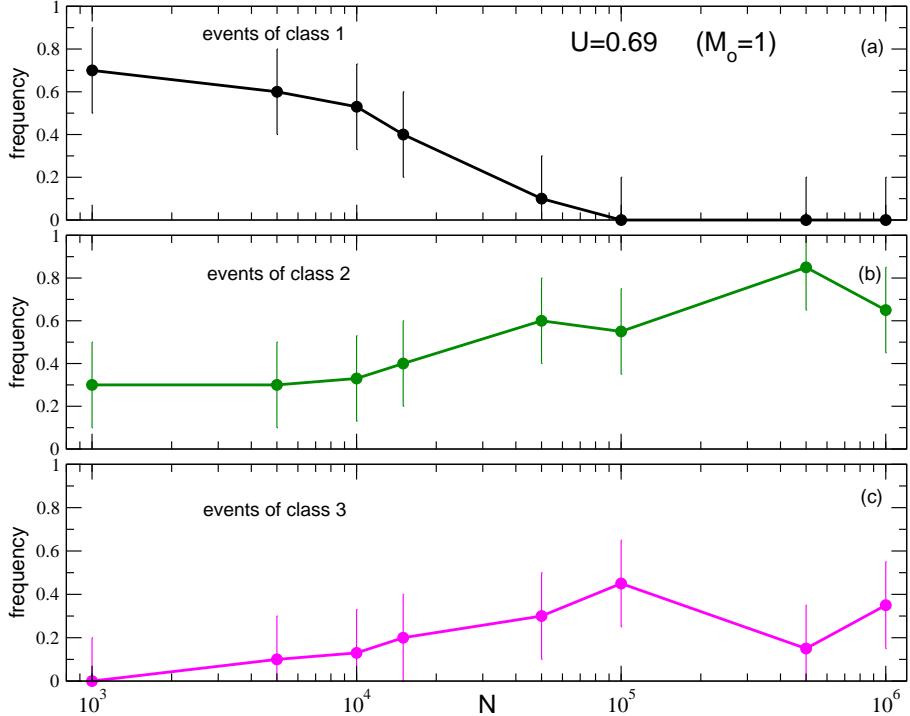}
\end{center}
\caption{ Relative frequency of occurrence for the three classes of events shown in Fig. 1 as a function of $N$.
A total of $m=20$ realizations for each $N$ was considered. We report also an uncertainty  equal to $\pm\Delta=1/\sqrt m$. The three curves add up to unity.}
\end{figure}

\section{Discussion of numerical results for the CLT }

In this section, following the prescription of the CLT and the procedure adopted in Refs. \cite{reply07,epl07,canberra}, we construct probability density functions (Pdf) of quantities expressed as a finite sum of stochastic variables and we select these variables along the deterministics time evolutions of the $N$ rotors. 
More formally, we study the Pdf of the quantity $y$ defined as
 \begin{equation}
  y_i=\frac{1}{\sqrt{n}}\sum_{k=1}^n  p_i(k)  ~~~ for  ~~~ i=1,2,...,N ~~,
\label{eq.3}
\end{equation}
where  $p_i(k)$, with $k=1,2,...,n$,  are the rotational velocities of the $ith$-rotor taken at fixed intervals of time $\delta$ along the same trajectory obtained integrating the HMF equations of motions. 
The product $\delta \times n$ gives the total simulation time over which the sum of Eq. (\ref{eq.3}) is calculated. 
As stressed in Ref. \cite{epl07}, the variables $y$'s are also proportional to the \textit{time average} of the velocities
along the single  rotor  trajectories (in fact the $1/\sqrt{n}$ scaling is not necessary, and has been adopted just to conform to usage).

\begin{figure}%[h,t]
\begin{center}
\epsfig{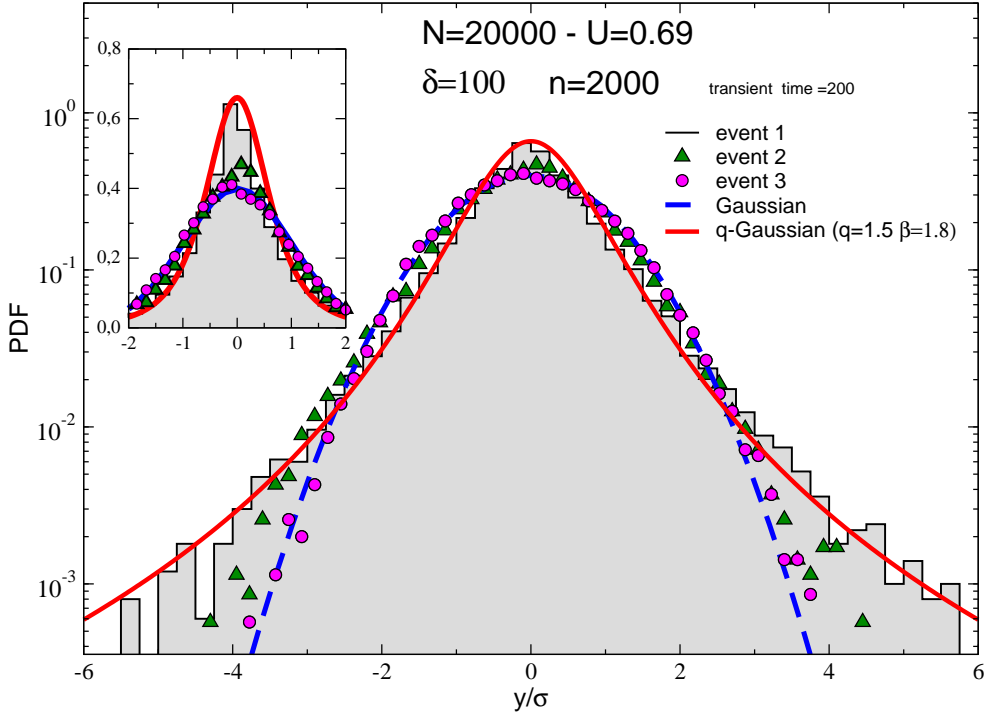}
\end{center}
\caption{ We  present  for each class of the  QSS  found, the   different  central limit theorem behavior observed. A Gaussian (dashed curve) with unitary variance and a  $q$-Gaussian  $p(x)=  A e_q(-\beta x^2)$ with $A=0.66$ $q=1.5$  and $\beta=1.8$  (full curve) are also reported for comparison. In the inset a  magnification of  the central  part  in linear scale is  plotted. 
See text for further details. 
}
\end{figure}

In Fig. 3 we show how the shape of the $y$'s Pdf is affected by the existence of different classes of events
in the QSS of the HMF model. We  consider   the same system of  the previous section with $N=20000$, $U=0.69$
and $M_0=1$ initial conditions. 
Then we plot the Pdfs corresponding to the three different 
events shown in Fig. 1, the sum (\ref{eq.3}) being  calculated over all the 
QSS extension with $\delta=100$ and $n=2000$ (for a total time of $200000$ steps, after a transient of $200$, see orizontal arrow in Fig.1). 
All the curves were normalized to their standard  deviations in order to compare them 
(notice  that, as done in \cite{reply07}, in the Eq. (\ref{eq.3}) we do not subtract an average value  calculated over the trajectory and all  rotors as in \cite{epl07,canberra} since it is not important  due to this  kind of normalization).
In case of no global correlations, as one would expect at equilibrium, the standard CLT holds and the resulting 
Pdf should be a Gaussian (reported as dashed line for comparison). This fact has been  verified for the equilibrium case in \cite{epl07}. Actually this is also  what happens  for  the
event 3 (full circles) and it is most  probably due to its similarity with the homogeneous QSS recalled 
in the previous section.       
On the other hand, if one considers the event 1 (gray histogram), strong deviations from the Gaussian behavior appears, 
meaning that global correlations become important. In this case a very good fit is obtained with  a $q$-Gaussian 
defined  as 
 \begin{equation}
  p(x)= A \, e_q(-\beta x^2)=\frac{A}{[1+(q-1)\beta x^2]^{1/(q-1)}} \,,
  \label{eq.4}
\end{equation}
where  $A(q,\beta)$ is a normalizing constant, which coincides in fact with $p(0)$.  An excellent fitting corresponds to $p(y/\sigma)$
$= 0.66 \, e_{1.5}^{-1.8 \,(y/ \sigma)^2}$ (full line), a result   very   similar to those  discussed  
in Refs. \cite{epl07,canberra}. 
Finally, the Pdf of event 2 (full triangles) lies between the previous two, being (within the time interval that has been observed) neither Gaussian nor $q$-Gaussian but
a sort of mixture of them, due to the presence of a double QSS plateau in the temperature evolution of this event.
This new scenario, where both Gaussian and $q$-Gaussian attractors appear in the QSS of a given system, implies that, if an average of the $y$'s is performed on many realizations, the particular mixture of the three kinds of events discussed here will affect the resulting Pdf, in particular for sizes of the system greater than $N=10000$ (i.e. when
the relative frequency of events of class 3 becomes important).
In this respect, such a scenario also qualifies the context within which the results presented in Refs. \cite{epl07,canberra} are to be considered, which is  thus not general, especially  for large system sizes.

\begin{figure}
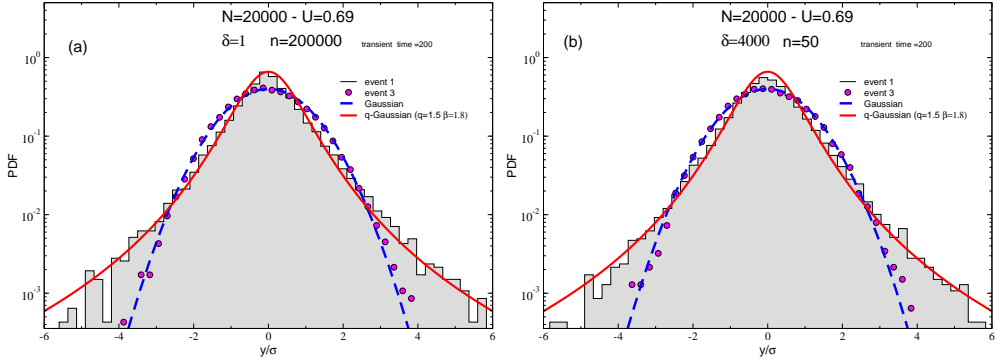
%[h,t]
\begin{center}
\epsfig{figure=fig4a.eps,width=6.5truecm,angle=0}
\epsfig{figure=fig4b.eps,width=6.5truecm,angle=0}
\end{center}
\caption{ CLT Pdfs for the same events 1 and 3 of Fig. 3, but with different values of $\delta$ and $n$. 
A Gaussian (dashed curve) with unitary variance and a  $q$-Gaussian with $A=0.66$, $q=1.5$  and $\beta=1.8$  (full curve) are also reported for comparison. 
}
\end{figure}

As a further remark, in Fig. 4 we test the robustness of the behavior observed in Fig. 3 for events 1 and 3.
We repeated  the calculations  on the same events of Fig. 3, but with a different   time interval $\delta$.
In the figure we report  the cases  $\delta=1$  (a) and $\delta=4000$ (b). 
One can see that changing $\delta$ and $n$ in the calculation of the $y$'s unaffects the Pdf's shape 
corresponding  to those events, providing that the total simulation time (i.e. the product $\delta \times n$)
remains unchanged. In this figure, the same Gaussian and $q$-Gaussian curves of Fig. 3 are plotted again for comparison.
It is important to notice  that, for the  case  $N=20000$ discussed in this article,   we  did not  study in detail the  behavior  of the CLT   for the  events  of class 2. In this case one could probably  distinguish 
   a first time evolution, closer   to events of class 1,  from a final  time evolution closer  to events of class  3.  In the  present case, however, the first part of the plateau is too short  to allow  the  investigation of the CLT behavior. The  latter  can probably be done    for  systems  of larger sizes: we leave this  study for the future. However, there is another  difficulty which  is intrinsic   with the events of this  class. In fact they  exhibit  a general  strong variability  as  shown   in Fig. 5, where we  plot the  temperature time evolution for several    different   events  of class 2   for the case   
$N=20000$. Although  the qualitative behavior with a  double plateau is the same,  the temperature  of the  initial plateau can vary. In this respect a  more detailed   investigation is in progress.

\begin{figure}%[h,t]
\begin{center}
\epsfig{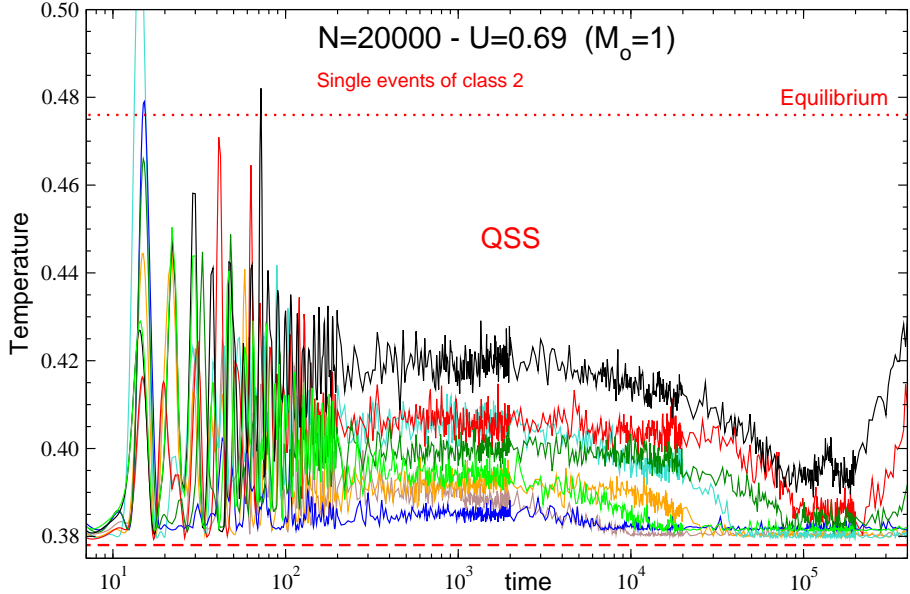}
\end{center}
\caption{ Different  events  of class  2 are plotted for the case N=20000.  A large variability is observed  for this class, at variance with the other two.}
\end{figure}

Before concluding, we  would like to illustrate that the  CLT behavior observed for the events of class 3 for   $M_0=1$ initial conditions is equivalent to that observed for any typical event with initial conditions $M_0=0$. In  Fig. 6  we 
plot the CLT behavior  for  $M_0=0$ in comparison with the case   $M_0=1$  for class 1 events, previously reported in Fig. 3.  The Gaussian Pdf obtained for the case  $M_0=0$ is  comparable  with what obtained  in Fig. 3
for the events of class 3. 

In Ref. \cite{epl07}, we  showed a  clear inequivalence between time averages and ensemble averages. In  Fig. 7, we show that this inequivalence persists also for different initial conditions and classes.  The ensemble averages of velocities are considered at time $t=1000$ inside  the QSS  over 10 events.  The Pdfs  plotted  in panels (a) and (b) are to be compared with the time averages curves shown in panel (a) of Fig. 6, while those plotted 
in panels (c) and (d) are to be compared  with the  time averages curves of events 1 and 3 of Fig. 3. A Gaussian (dashed curve) with unitary variance and a  $q$-Gaussian with $A=0.66$, $q=1.5$  and $\beta=1.8$  (full curve) are also reported for comparison. It is clear that 
the curves in Fig. 7 are neither  Gaussian nor q-Gaussians.

\begin{figure}   %[h,t]
\begin{center}
\epsfig{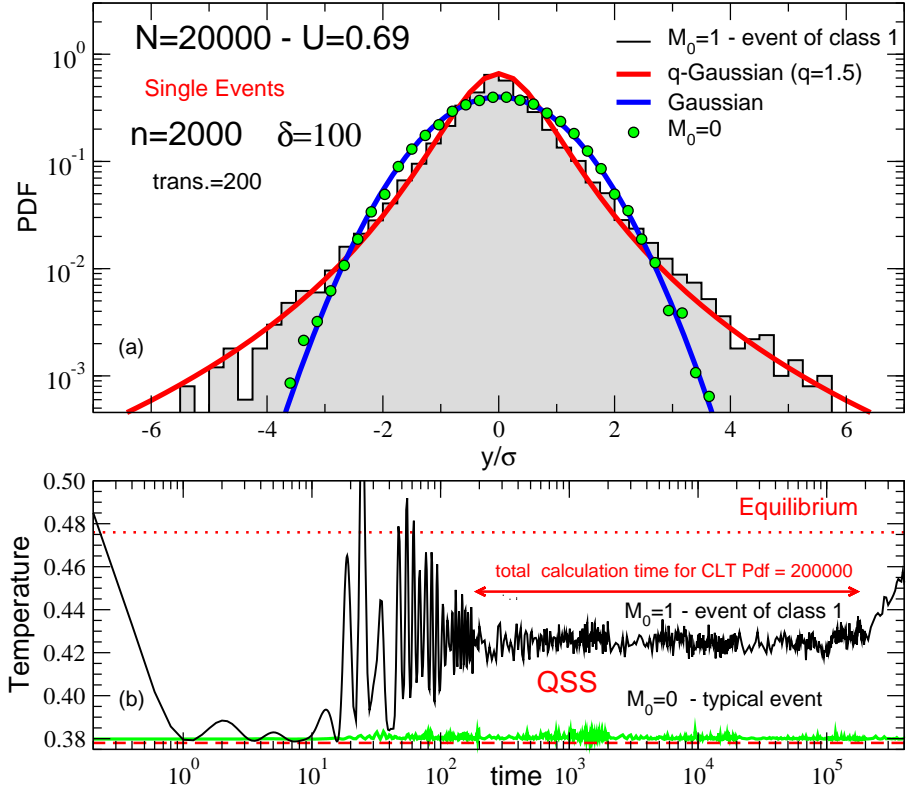}
\end{center}
\caption{ (a) Comparison of the CLT behavior for the case $U=0.69$  initial magnetization $M_0=1$ (class 1)  vs  $M_0=0$. 
The size of the system is $N=20000$. A Gaussian (dashed curve) with unitary variance and a  $q$-Gaussian with $A=0.66$, $q=1.5$  and $\beta=1.8$  (full curve) are also reported for comparison. 
(b) Temperature  time evolutions of the same events shown in panel (a).
}
\end{figure}

\begin{figure}   %[h,t]
\begin{center}
\epsfig{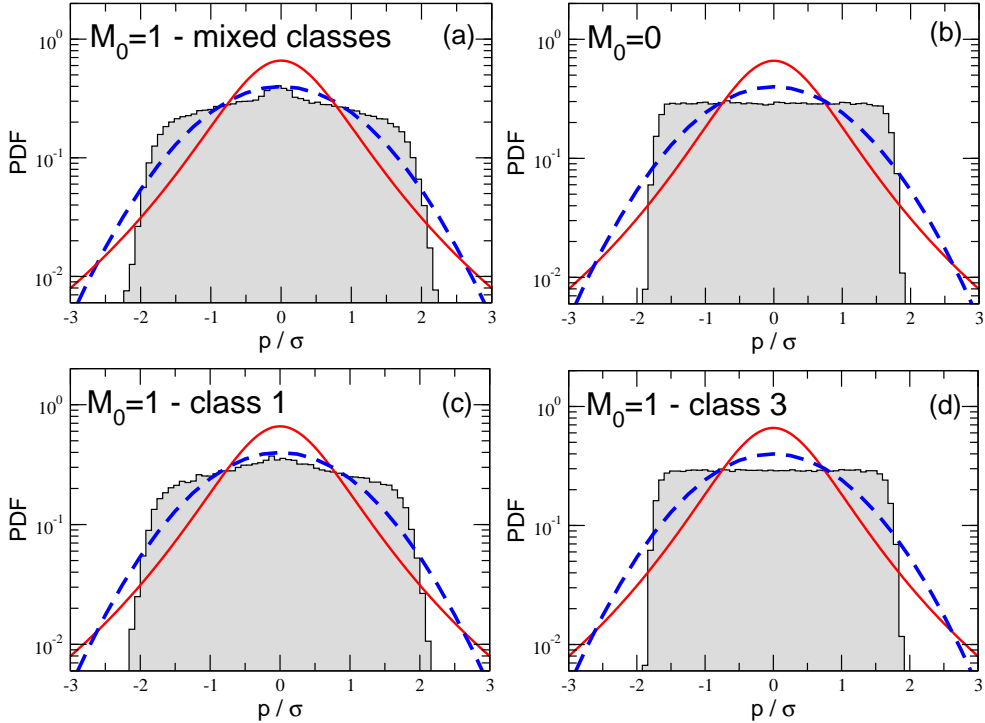}
\end{center}
\caption{ Ensemble  averages of velocities for $N=20000$, $U=0.69$  taken at time $t=1000$ in the QSS regime for different initial conditions and classes of events. The average was done over 10 events.  A Gaussian (dashed curve) with unitary variance and a  $q$-Gaussian with $A=0.66$, $q=1.5$  and $\beta=1.8$  (full curve) are also reported for comparison. See text for further details.}
\end{figure}

\section{Conclusions}

On the  basis  of the new calculations presented here, one should distinguish among different 
classes  of QSS for a given size  of  the HMF system. Our tests do confirm that correlations  
can be different for different dynamical realizations of the same system, starting from the same 
class of initial conditions, and therefore also the  Central Limit  Theorem behavior  can change.   According to 
the  class  considered  we can have a Gaussian Pdf, a $q$-Gaussian one  or a mixture  between  the two.
In this respect, the presence of  a $q$-Gaussian curve  confirms the results  previously published about the indications  for a generalized  CLT in long-range  Hamiltonian systems  when ergodicity is broken  and correlations  are strong enough.
The present results, where we verify that the time evolution of the system is quite sensitive to the specific initial conditions, reminds what occurs in other anomalous systems, like the Hydrogen atom as studied  recently in 
Refs. \cite{Oliveira1,Oliveira2}. 
We  have  checked   also that inequivalence between time averages and ensemble  averages continues to hold
within the different  classes  for $M_0=1$ and  for $M_0=0$ initial conditions.
Finally, although  it is true that   the frequency of occurrence of 
events of class 1 and 2  for very large  sizes, and 
in particular  the  attractor  for the events of class  2,   
 remain to be investigated  with more accuracy  (calculations  with  better   statistics are in progress)  before advancing definitive conclusions,  from these  simulations  one could  conjecture  that   the events  of class 1 tend to  disappear  in the thermodynamic limit. However, even if this  was the case, their probability of occurrence  (as well as the probability of occurrence of the first metastable plateau of events of class 2) can be significantly different from zero for  large  but finite systems and therefore  the  applicability  
 of $q$-statistics in metastable states of real complex systems remains a valid and interesting possibility.

\section*{Acknowledgments}

The  present calculations  were done within the Trigrid project   and we thank   M. Iacono Manno for technical help.  
A.P. and A.R. acknowledge
financial support from the PRIN05-MIUR project "Dynamics and
Thermodynamics of Systems with Long-Range Interactions". C.T. acknowledges partial financial support from the Brazilian Agencies CNPq and Faperj.

\end{document}